\documentclass [fleqn,12pt]{article}

\usepackage{amsmath,amssymb,latexsym,cite}
\usepackage[english]{babel}
\usepackage{graphicx}
\usepackage{epstopdf}
\usepackage{caption}
\usepackage{subcaption}
\usepackage[T2A]{fontenc}
\usepackage[utf8]{inputenc}
\def\nqq{\hspace*{-2em}}

\def\cm{\hspace*{1cm}}
\def\inch{\hspace*{1in}}
\def\ten#1{\mbox{$\, \cdot\, 10^{#1}$}}

\def\lal{&&\nqq {}}
\def\eq{Eq.\,}
\def\eqs{Eqs.\,}
\def\beq{\begin{equation}}
\def\eeq{\end{equation}}
\def\bear{\begin{eqnarray}}
\def\bearr{\begin{eqnarray} \lal}
\def\ear{\end{eqnarray}}
\def\earn{\nonumber \end{eqnarray}}

\def\nnn{\nonumber\\ \lal }
\def\nnnv{\nonumber\\[5pt] \lal }

\def\yyy{\\[5pt] \lal }

\def\tst{\textstyle}

\def\fract#1#2{{\tst\frac{#1}{#2}}}

\def\half{{\fract{1}{2}}}

\def\qua{\fract 14}
\def\e{{\,\rm e}}
\def\d{\partial}

\def\sign{\mathop{\rm sign}\nolimits}
\def\diag{\mathop{\rm diag}\nolimits}

\def\const{{\rm const}}

\def\ep{\epsilon}
\def\then{\ \Rightarrow\ }

\usepackage{color}
\def\red#1{{\color{red} #1}}
\def\blue#1{{\color{blue} #1}}
\def\eqn#1{\eq\eqref{#1}}
\def\rf{\eqref}
\def\mn{_{\mu\nu}}

\def\mN{_\mu^\nu}

\def\kappa{\varkappa}
\def\ep{\epsilon}

\def\og{\overline{g}{}}

\def\ssph{static, spherically symmetric}

\begin{document}

\title{Inhomogeneous compact extra dimensions}

\author{K.~A.~Bronnikov$^a$\thanks{kb20@yandex.ru}, R.~I.~Budaev$^b$, A.~V.~Grobov$^{b,e}$\thanks{alexey.grobov@gmail.com}, \\
	A.~E.~Dmitriev$^b$\thanks{alexdintras@mail.ru}, Sergey G. Rubin$^{b,c}$\thanks{sergeirubin@list.ru} 
}

\date{\small\it
	\vspace{4mm}
	\begin{description}
		\item[$^a$] Center of Gravity and Fundamental Metrology, 
		VNIIMS, 46 Ozyornaya st., Moscow 119361, Russia;\\
		\item[$^b$] National Research Nuclear University MEPhI 
		(Moscow Engineering Physics Institute), 115409 Moscow, Russia;\\
		\item[$^c$] Institute of Gravitation and Cosmology, 
		Peoples' Friendship University of Russia (RUDN University),
		6 Miklukho-Maklaya St., Moscow 117198, Russia;\\
		\item[$^d$] N.I. Lobachevsky Institute of Mathematics and Mechanics,
		Kazan  Federal  University, 
		Kremlevskaya  street  18,  420008  Kazan,  Russia\\
		\item[$^e$] National Research Center "Kurchatov Institute", 
		Akademika Kurchatova pl., 1, 123182 Moscow, Russia\\
	\end{description}    
}
\maketitle

\vspace{-4mm}
\begin{abstract}
	We show that an inhomogeneous compact extra space possesses two necessary features ---
	their existence does not contradict the observable value of the cosmological constant
	$\Lambda_4$ in pure $f(R)$ theory, 
	and the extra dimensions are stable relative to the ``radion mode'' 
	of perturbations, the only mode considered.
	For a two-dimensional extra space, both analytical and numerical solutions for the metric 
	are found, able to provide a zero or arbitrarily small $\Lambda_4$.
	A no-go theorem has also been proved, that maximally symmetric compact extra spaces are 
	inconsistent with 4D Minkowski space in the framework of pure $f(R)$ gravity.   
\end{abstract}

\section{\label{sec:introduction} Introduction}

In modern physics, the idea of extra dimensions is used for explanation of a variety 
of phenomena. This idea is applied for elaboration of physics beyond the Standard
Model, cosmological scenarios including inflationary models and the origin of the dark
component of the Universe (dark matter and energy), for explaining its baryon asymmetry, 
the number of fermion generations and so on. Gradually, it is becoming
the main element for a future theory. 

With their whole diversity, the models involving extra dimensions must overcome 
two common problems: the smallness of the cosmological constant they induce 
\cite{Lambdasmall1, Lambdasmall2}, and the stability of an extra space metric.
Observations indicate that our Universe is nowadays expanding with acceleration
\cite{accel}, which can be explained in Einstein gravity by the existence of a
cosmological constant $\Lambda \sim 10^{-3}~{\rm eV}^4$ or something very close 
to it, called dark energy, see a review in \cite{StarL}. We live in space-time
approximately described by the de Sitter metric, which is, in a sense, extremely 
close to the Minkowski one. Any study should be in agreement with this 
observational fact.

There are a lot of approaches to solving the problem: one introduces scalar fields
(see, e.g., \cite{Carugno}), form fields \cite{Hawk}, invariants other than 
powers of $R$ \cite{Ru16}, or considers compact extra spaces with zero curvature,
e.g., Calabi-Yau manifolds \cite{Zhuk13}, one-dimensional extra spaces 
\cite{Ponce} and flat tori \cite{Flatorus}. As shown in \cite{Navarro}, 
insertion of form fields leads to an extra space with the metric of a sphere 
with a deficit angle and an observationally acceptable value of $\Lambda$. The study appeared recently is based on Horndeski scalar-tensor theories 
of gravity involving Poincare invariance breaking \cite{FabFour} that comes
together with a self-tuning mechanism driven by scalar field dynamics. It was
pointed out by Babichev and Esposito-Farese that a local deviation from general
relativity would rule out a model even with good $\Lambda$ tuning. The accordance
of Hordeski-like theories (generalized Gallileon) with solar system gravity 
tests were studied in \cite{localtest}.

On the other hand, inhomogeneous extra spaces are a promising tool that 
could underlie such effects of low-energy physics as the formation of fermion
generations \cite{Gogber}, explanation of dark matter \cite{RDG}, and creation 
of branes \cite{RuLocali}.  
Evidently, this direction deserves a deep study. In particular, both problems
mentioned above, stability of compact extra dimensions and smallness of the
cosmological constant, should be solved at least at the classical level. 
In this paper we  discuss these problems for an inhomogeneous (deformed) compact
extra space with a nonzero Ricci scalar, related to \cite{RDG, RuLocali}. 

This study is performed on the basis of pure multidimensional $f(R)$ gravity. 
No other tools listed above are used. The interest in $f(R)$ theories is motivated
by inflationary scenarios starting from Starobinsky's pioneering work
\cite{Starf}. A number of viable $f(R)$ models in 4D space that satisfy the
observational constraints have been proposed in 
\cite{Amendola,Starobinsky,Cognola,Criscienzo}. At first glance, 
it seems to be a hard task to obtain a zero or extremely small curvature of our 
4D space if the Ricci scalar of extra space is not negligibly small. 
A reason is that the Ricci tensor of the extra space contributes
to the 4D part of the multidmensional field equations, and therefore 4D flat 
or weakly curved 4D space-times fail to be solutions of these equations; see 
some details in Section 3 where we prove that if the extra space is a direct
product of maximally symmetric manifolds, the 4D Minkowski metric cannot appear 
in the solutions, be it in the Einstein or Jordan conformal frame.

Nevertheless, we show that the problem can be solved if the extra space 
and hence its curvature is inhomogeneous. We discuss the ways to construct
a set of ``almost Minkowski 4D spaces". In what follows, except for 
Section 6, we omit the word "almost" and simply mention flat Minkowski space 
as a representative of the set.

A separate problem is that of stability of the extra space geometry
\cite{Green,Carroll,Nasri}. A static solution can be obtained by 
taking into account the Casimir effect \cite{Candelas} or, for example, 
form fields \cite{Freund,Kubyshin}. Another possibility was discussed 
in \cite{Yoshimura,Svadk}: it was shown that if the scale factor $a(t)$
of our 3D space is much larger than the scale factor $b(t)$ of the extra
dimensions, a contradiction with observations can be avoided. Section 6 
of the present paper is devoted to the stability problem for our deformed 
extra space models.



\section{Basic equations}

Consider, in $D$-dimensional space-time with the metric $g_{AB}$, a theory 
of gravity with higher-order derivatives and the action in the form
\beq          \label{act1}
S=\frac{m_D ^{D-2}}{2}\int d^D x \sqrt{g_D}[f(R) ]
\eeq
with an arbitrary function $f$ of the the D-dimensional scalar curvature 
$R = R_D$ and the D-dimensional Planck mass $m_D$; $g_D = |\det g_{AB}|$,
and the indexes $A,B, \ldots$ refer to all $D$ dimensions. In the following  
consideration we choose the units in which $m_D = 1$.

The corresponding equations of motion read
\beq                    \label{eq-gen}
-\frac{1}{2}f(R)g_{AB}
+ (R_{AB} + \nabla_{A}\nabla_{B} - g_{AB}\square_D) f_R = 0,
\eeq
where $\square_D = \nabla^A \nabla_A$, $f_R \equiv df/dR$.
Assuming $D=n+4$, we will consider metrics of the form
\beq                                 \label{ds1}
ds^2 = g_{AB} dx^A dx^B = g\mn(x) dx^\mu dx^\nu + g_{ab}(x,y) dy^a dy^b,
\eeq
where the indexes $\mu, \nu, \ldots$ refer to the 4D part of the metric, and  
$a,b,\ldots$ to its extra-dimensional part; $(x)$ and $(y)$ mark the 
dependence on $x^\mu$ and $y^a$, respectively.

Notice that the $(\mu,\nu)$ components of the metric tensor are assumed 
to be independent of the extra coordinates $y^a$. This leads to some  
integro-differential equations for $g_{AB}$ instead of the equations
following directly from \eqref{eq-gen}, see a discussion after 
\eq \eqref{ab-trace}.

A substantial simplification of the field equations is achieved if we 
suppose that our metric varies slowly along the 4D coordinates of our 
space-time as compared to the extra coordinates $y^a$. More specifically,
\beq          						\label{ep}
\partial_{\mu} g_{AB} \sim \epsilon \partial_a  g_{AB},
\eeq
that is, each $\partial_\mu$ contains a small parameter $\epsilon$.

The general expression for the scalar curvature of space-time with 
the metric \rf{ds1} is
\bearr                \label{R_D} 
R = R_4 (x) + R_n (x,y) + f_X (x,y),
\nnnv
f_X  \equiv -\nabla_{\mu} X^{\mu}  + \qua X_{\mu} X^{\mu}
- \qua (\partial^{\mu} g^{ab}) (\partial_{\mu} g_{ab}),
\cm
X^{\mu} \equiv g^{ab} \partial^{\mu} g_{ab},
\ear
where $R_4 = R_4[g\mn] = R^\mu_\mu$, $R_n = R_n[g_{ab}] = R^a_a$ are the 
scalar curvatures of the corresponding subspaces, and $\nabla_{\mu}$ is 
the covariant derivative in the metric $g\mn$. It follows from \eqref{ep} 
that the 4D space is very weakly curved as compared to the extra dimensions,
that is, the Ricci tensor $R\mN = O(\ep^2)$, and, since $X^\mu = O(\ep)$,
\beq
R \equiv R^A_A = R_n + O(\ep^2).       \label{ep-R}
\eeq

Under the condition \eqref{ep}, the Taylor expansion of $f(R)$ in 
Eq.\,(\ref{act1}) gives
\bearr   \label{act2}
S= \frac{1}{2}\int d^4 x d^n y \sqrt{g_4 g_n}~ f(R_4 + R_n + f_X)
\nnn \quad
\simeq \frac{1}{2}\int d^4 x d^n y\sqrt{g_4 g_n}
[  f(R_n) + f_R(R_n) (R_4 + f_X)  + o(\ep^2)],
\ear

In the next section we show that maximally symmetric extra spaces restrict our abilities in the description of the modern stage of our Universe.

\section{A maximally symmetric extra space is incompatible with 4D flat space}

In this section we prove a \textit{no go theorem: in the framework of pure $f(R)$ gravity, 4D Minkowski space-time is incompatible with compact, maximally symmetric extra spaces of nonzero curvature.}

The metric of our 4D space is extremely weakly curved, i.e., almost flat, as compared with possible extra-dimensional curvatures. This theorem 
shows that to be in agreement with the modern state of the Universe,
a compact extra space must either have zero curvature or not be 
maximally symmetric, if not accompanied by other fields. In this paper 
we consider in detail a compact inhomogeneous extra space with 
nonzero curvature. It is its inhomogeneity that leads to a small 
or zero cosmological constant without violating the theorem.

To prove the theorem, let us analyze the expression \eqref{act2} with 
the metric \eqref{ds1} where
\begin{equation}     \label{eq_8}
g_{ab}(x,y) = e^{2\beta(x)} \og_{ab}(y),
\end{equation}
and $\og_{ab}(y)$ describes a space of constant nonzero curvature.

A few words to justify this ansatz. 
The characteristic size $l$ of a compact extra space is smaller than 
$10^{-18}$ cm, or the energy scale is higher than 10 TeV. Excitations of a 
compact extra space geometry are known to form a Kaluza-Klein tower with 
energies $E > l^{-1}$. The energies we deal with are many orders of magnitude 
smaller, therefore excitations are suppressed, and the extra space metric 
$g_{ab}$ represents a stationary configuration described by the $(ab)$ part of 
the classical equations of motion and a uniform distribution in our space. 
Hence, if we decompose the metric $g_{ab}(x,y)$ as
\[
g_{ab}(x,y) = \sum_k f_n(x)h_{ab}^{(k)}(y)
\]
where $h_{ab}^{(k)}(y)$ is an orthonormal set of solutions to the corresponding 
equations of motion, there remains only the first term with $k=1$, and we 
arrive at Eq.\,\eqref{eq_8}. The latter leads to the following expressions
\bearr
\sqrt{g_n(x,y)} = e^{n\beta(x)}\sqrt{\og_n(y)},
\yyy                    \label{phi}
R_n(\beta(x),y) = \bar{R}_n \cdot e^{-2\beta(x)} \equiv \phi(x),
\ear
where $\bar{R}_n \ne 0$ is a constant.
The expression \eqref{act2} can be rewritten as
\begin{multline} \label{act3}
S \simeq \frac{1}{2}\int d^4 x d^n y\sqrt{g_4 g_n}
[  f(R_n) + f_R(R_n) (R_4 + f_X)  + o(\ep^2)] 
\\
= \frac{m_4^2}{2} \int d^4 x\sqrt{g_4}e^{n\beta(x)} 
[ f(\phi) + f_R(\phi) (R_4 + f_X) + o(\ep^2)],  \inch
\end{multline}
where
\begin{equation}
m_4^2 = \int d^n y\sqrt{\bar{g}_n}, \qquad 
g_n(x,y) = e^{2n\beta(x)}\bar{g}_n(y).
\end{equation}

The expression \rf{act3} represents a 4D scalar-tensor theory of gravity with
the metric $g\mn$ and the scalar field $\beta$ (or $\phi$), specified in a 
Jordan conformal frame. We want to obtain a constant equilibrium value of
$\beta$, which can be achieved after a transition to the Einstein frame in which 
the scalar field dynamics is separated from that of the metric. This is 
done using the conformal mapping \cite{BlackHoles}
\begin{equation}
\tilde{g}_{\mu\nu}(x) = |Y(\phi(x))|g_{\mu\nu}(x), \qquad 
Y(\phi(x)) \equiv e^{n\beta(x)} f_R(\phi(x)).
\end{equation}
Now the action acquires the form \cite{BlackHoles}
\begin{equation} \label{ScalarAction}
S\simeq\frac{m_4^2}{2} \int d^4 x\sqrt{\tilde{g}_4} 
[ (\sign\,F_R(\phi)) (\tilde{R}_4 + K(\phi) - 2W(\phi)],
\end{equation}
where $K(\phi)$ is a kinetic term of the field $\phi(x)$, and
\begin{equation}
-2 W(\phi) = e^{-n\beta}\frac{f(\phi)}{f_R(\phi)^2} 
= \left(\frac{\phi}{\bar{R}_n}\right)^{n/2} \frac{f(\phi)}{f_R(\phi)^2},
\end{equation}
so that $W(\phi)$ is the potential energy density of $\phi$. 

We intend to consider the physical situation in the modern epoch, such 
that the scalar field $\phi(x)$ corresponds to a uniform distribution of 
dark energy, which is almost constant. Therefore we assume that $\phi$ has 
been settled at an extremum of its potential.\footnote
{Depending on the particular form of $f(R)$, the kinetic term 
	$K(\phi)$ can have any sign. If it is positive, we are dealing with a
	usual, canonical scalar field, and its stable equilibrium corresponds to
	a minimum of $W$. If $K(\phi) <0$, the field is phantom and has a 
	stable equilibrium at a maximum of $W$. Thus it is more precise
	to speak of an extremum.
}
Let there be an extremum of $W(\phi)$ at some $\phi=\phi_c$.
To get flat 4D space, we should require
\beq		\label{zeroLa}
W(\phi_c) = 0;\qquad \frac{dW}{d\phi} (\phi_c) = 0.
\eeq
This gives
\bearr \label{w0}
|\phi_c|^{n/2}\frac{f(\phi_c)}{f_R(\phi_c)^2} = 0,
\yyy       \label{w'0}
\left[\frac{n}{2\phi}\frac{f}{f_R^2}+ \frac 1{f_R} 
- \frac{2f f_{RR}}{f_R^3} \right]_{\phi = \phi_c} = 0,
\ear
where $f_{RR}\equiv d^2 f(R)/dR^2$. Note that $\phi\neq 0$ due to \eqref{phi}, 
and since $\bar{R}_n\neq 0$, \eq \eqref{w0} gives $f(\phi_c)=0$, 
then \rf{w'0} gives $f_R(\phi_c)=\infty$, which is unacceptable.
This completes the proof. 

A question arises: can it happen that, unlike a precisely flat 4D space-time, a weakly curved, say, de Sitter space-time can still be compatible with a
homogeneous extra space? In other words, does the above theorem cover not   only precisely Minkowski but also weakly curved 4D space-times?

The answer depends on how small is the deviation from the flat metric. 
The actually observed 4D curvature is very small: the cosmological almost 
dS metric has a curvature radius around $10^{28}$ cm, enormously huge as 
compared to the length scale ($\lesssim 10^{-18}$ cm) of the extra space,
making the curvature smaller by a factor of $\gtrsim 10^{92}$.
Even if one tries to obtain, by choosing the form of $f(R)$ and other 
parameters, the cosmologically observed value of 
$\Lambda \sim 10^{-123} m_{\rm Pl}^2$, where $m_{\rm Pl}$ is the Planck mass, 
then, in any case, the calculations are performed with some finite accuracy 
with errors manifestly larger than $10^{-123}$, therefore such a choice will 
not be distinguishable from that leading to $\Lambda =0$. The lesson is that 
it does not make sense to lose time for such attempts, and one should instead
try to involve new ingredients like matter fields or deformed extra spaces.

The above theorem also holds for a direct product of any number of 
maximally symmetric compact extra spaces. In this case, in \rf{ds1}
we have 
\beq
g_{ab}(x,y) dy^a dy^b = \sum_{i=1}^{N} e^{2\beta_i(x)} \og^{(i)}, 
\eeq  
where $N$ is the number of $d_i$-dimensional extra factor spaces of 
constant curvatures 
$\bar{R}_i \ne 0$, with $x$-independent metrics $\og^{(i)}$. Then,
\bearr
R = R_4 (x) + \sum_i R_i (x,y) + f_X (x,y) 
= R_4 (x) + \sum_i e^{-2\beta_i(x)} \bar{R}_i (y) + f_X (x,y),
\yyy
R_i (\beta(x),y) = \bar{R}_i \cdot e^{-2\beta_i(x)} \equiv \phi_i(x),
\qquad  \phi(x)\equiv \sum_i \phi_i(x),
\yyy
f(R) = f(\phi) + f_R(\phi) ( R_4 + f_X ) + o(\ep^2),
\yyy
m_4^2 = m_D^{D-2}
\int d^{d_1} y d^{d_2} y...\sqrt{\bar{g}_1 \bar{g}_2 ...},
\yyy
-2W(\phi) = \exp {\Big(-\sum_i d_i\beta_i}\Big)
\frac{f(\phi)}{f_R(\phi)^2} 
= \left(\frac{\phi_1}{\bar{R}_{n_1}}\right)^{d_1/2} 
\left(\frac{\phi_2}{\bar{R}_{n_2}}\right)^{d_2/2} ... \cdot 
\frac{f(\phi)}{f_R(\phi)^2}.
\ear
The condition \eqref{zeroLa} now turns into
\begin{equation}
W(\phi_{1c},\phi_{2c},...) = 0, \qquad 
\d W/\d \phi_i (\phi_{1c},\phi_{2c},...) = 0, \qquad i=1,2,...N,
\end{equation}
where $\phi_{i}=\phi_{ic}$ is the assumed extremum of $W$. This gives:
\bearr   \label{wi0}
\left(\frac{\phi_{1c}}{\bar{R}_1}\right)^{d_1/2} 
\left(\frac{\phi_{2c}}{\bar{R}_2}\right)^{d_2/2} ... \cdot 
\frac{f(\phi_c)}{f_R(\phi_c)^2} = 0,
\yyy       \label{wi'0}
\left[\frac{d_i}{2\phi_i}\frac{f}{f_R^2}+ \frac 1{f_R} 
- \frac{2f f_{RR}}{f_R^3} \right]_{\phi_i = \phi_{ic}} = 0
\ear
with $i = 1,2,...N$. As above, it is easy to see that \eqref{wi0} leads 
to $f(\phi_c)=0$ and $f_R(\phi_c) = \infty$, which is unacceptable. 
Therefore flat 4D space cannot be obtained if the extra spaces are 
maximally symmetric.

Another feature of interest follows from the condition \eqref{wi'0} 
in the case with a nonzero extremum of $W$:
$W(\beta_{1c},\beta_{2c},...) = \const\ne 0$, so that $f(R)\ne 0$ 
at $\phi_i = \phi_{ic}$. One has
\beq         \label{cond}
\frac{\phi_{ic}}{d_i} = \frac{f f_R}{2[ 2 f f_{RR} - f_R^2]}
\,\biggr|_{\phi_k = \phi_{kc}}, \qquad i,k = 1,2,...N.
\eeq
This equation is correct for any sign of $\phi_i$. (Note that
$\sign \phi_i = \sign\bar{R}_i$, the curvature sign of a factor spaces. )

Immediate consequences of the expressions \eqref{cond} are:
\begin{enumerate}
	\item 
	At an extremum of the potential, there are simple relations between the Ricci 
	scalars of the extra factor spaces:
	\beq
	R_{n_i}/R_{n_k} = d_i/ d_k,  \quad i, k = 1,2,...N. 
	\eeq
	\item 
	The Ricci scalars of all extra spaces should have the same sign for $W$ 
	to have an extremum. 
\end{enumerate}

All this was obtained in the Einstein frame in four dimensions. However, 
at constant $\beta_i$ the conformal factor $Y$ between the two frames is constant,
therefore all conditions obtained for an equilibrium state of $\beta_i$ are
valid if we work in the Jordan frame as well.

One can also note that one cannot exclude that, in addition to the extra 
factor spaces with nonzero curvature, there are some others, with 
$\bar{R}_i =0$ (e.g., toroidal ones). Such factor spaces simply 
do not take part in the above relations, which thus remain unaffected. 

\section{Zero-order approximation. Analytical treatment}

Let us now return to configurations with a single but deformed 
extra factor space in order to show that this concept can open new ways 
in the description of our universe evolution.
In what follows we restrict ourselves to static extra space-times
and impose certain boundary conditions determining the structure
of extra dimensions. We also use the slow-change approximation
\rf{ep}.

If the extra-space metric $g_{ab}$ is $x^\mu$-independent, 
(that is, $\beta (x)=0$ without loss of generality),
then in our 4D space-time we obtain general relativity with
a cosmological constant. Indeed, a comparison of the second 
line in \eqref{act2} with the Einstein-Hilbert action
\beq \label{HilbEin}
S_{\rm EH}=\frac{m_4^2}{2} \int d^4x  \sqrt{|g(x)|}(R_4 -2\Lambda_4)
\eeq
gives the Planck mass and the cosmological constant:
\bearr                \label{Lam}
m_4^2 = m_D^{D-2} \int d^n y \sqrt{g_n(y)}f_R(R_n),\qquad
\Lambda_4 =  -\frac{1}{2m_4^2}\int d^n y \sqrt{g_n(y)}\,f(R_n).
\ear
both depending on the static geometry $g_{ab}(y)$.

\subsection{General consideration}

In the order of magnitude $O(1)$ in terms of the small parameter $\ep$, 
the curvature of 4D space-time is negligible. Therefore, assuming for 
simplicity that there are only two extra dimensions, we consider the 
following metric with, in general, inhomogeneous curved compact extra space:
\beq          \label{ds_6}
ds^2 = g\mn dx^\mu dx^\nu + g_{ab} dy^a dy^b \equiv
\eta\mn dx^\mu dx^\nu  - e^{2\lambda}(dy^2  + dz^2),
\eeq
where $\eta\mn = \diag(1, -1, -1, -1)$ is the Minkowski metric, while $y$ and
$z$ are extra-space coordinates; the 2D metric $g_{ab}$ is chosen in a conformally
flat form without loss of generality, but we assume that $\lambda$ is a function of $y$ only.


In the previous section we have shown that with a maximally symmetric compact 
extra space the Minlowski 4D geometry is impossible. This, however, does not
exclude other geometries, which can, in particular, have the form of a 
deformed 2-sphere, but in any case the requirement $\Lambda_4 =0$ should 
be satisfied. (Let us note for clarity that an ordinary 2-sphere 
of radius $r_0$ is described in \rf{ds_6} by the function 
$\e^\lambda = r_0/\cosh y$, 
and the usual angular coordinates $\theta$ and $\varphi$ are connected with $y$ and $z$ 
by the transformation $1/\cosh y = \sin \theta$, $z=\varphi$.)

Let us try to find such solutions.
The $y$ dependence of the 2D metric to be found means that it is in general not
maximally symmetric but is deformed in some way. In the present approximation, 
nonzero are only the diagonal Ricci tensor components $R^a_b$.
The 6-dimensional Ricci tensor and scalar are expressed in terms of $\lambda(y)$ as
(the prime means $d/dy$)
\beq          \label{R_2}
R^y_y  = R^z_z = - e^{-2\lambda}\lambda'', \qquad
R = R_2 = -2 e^{-2\lambda}\lambda''.
\eeq
The ${a \choose b}$ components of (\ref{eq-gen}) read
\beq           							\label{ab}
- \delta^a_b f(R)/2 + (R^a_b + \nabla^a \nabla_b - \delta^a_b \Box_6)f_R (R) = 0,
\eeq
where $\Box_6 f_R = \Box_2 f_R = \nabla^a \nabla_a f_R$.
The trace of (\ref{ab}) gives
\beq                                        		    \label{ab-trace}
\Box f_R = -f + R^a_a f_R.
\eeq

The mixed ${\mu \choose a}$ components of (\ref{eq-gen}) are trivial.
As to the ${\mu \choose \nu}$ components,  there is a subtle point: instead of directly 
writing these components of \eqref{eq-gen},  it makes sense to return to the variation 
procedure taking into account the independence of $g\mn$ of $y^a$. Then, instead of 
the $({}\mN)$ components of \eqref{eq-gen}, variation of the action in $g\mn$ leads to
\beq  \label{mn-int0}
\int \sqrt{g_2[y]} d^2 y [f(R) + 2\Box f_R] =0,
\eeq  
if we assume the Minkowski 4D metric. Taking into account (\ref{ab-trace}), this can be transformed to
\beq                                             \label{mn-int}
\int \sqrt{g_2[y]} d^2 y [2R f_R -f(R)] =0,
\eeq
where (recall) $R = R^a_a$. Thus in this approximation we must solve  (\ref{ab}) and
apply the integral condition (\ref{mn-int}) to the solutions obtained.

More than that, the condition $\Lambda_4 =0$ with \eqn{Lam} together with \rf{mn-int}
imply that, separately,
\beq                                             \label{int-0}
\int \sqrt{g_2[y]} f(R)\, d^2 y=0, \qquad 
\int \sqrt{g_2[y]} R\,f_R\,d^2 y=0.
\eeq

There are two noncoinciding equations in (\ref{ab}), and their difference reads
\beq                                                          \label{vv1}
(\nabla^y\nabla_y - \nabla^z\nabla_z) f_R =0
\ \ \then \ \
\Box f_R =  2 \nabla^z\nabla_z f_R = -2e^{-2\lambda}\lambda' (f_R)'
\eeq
Writing this difference in its explicit form and integrating, we have
\beq                                                          \label{vv2}
(f_R)'' = 2\lambda' (f_R)' \ \ \then \ \
(f_R)' = A_0 e^{2\lambda}, \cm   A_0 = \const,
\eeq
and a further substitution into the trace equation \rf{ab-trace} gives a
second-order equation with respect to $\lambda(y)$:
\beq
f(R) - R f_R  = 2A_0 \lambda'.                 \label{eq-2o}
\eeq

We had two components of \eqs \rf{ab}, or, equivalently, their sum 
\rf{ab-trace} and difference \rf{vv1}, which are of fourth order with respect 
to $\lambda(y)$. Since there is only one unknown function 
$\lambda(y)$, these equations must not be independent, otherwise the 
system will be overdetermined. Let us show that both \rf{ab-trace} and 
\rf{vv1} follow from the single second-order equation \rf{eq-2o}.  

First, applying $d/dy$ to \rf{eq-2o} and expressing $\lambda''$ 
in terms of $R$ using \rf{R_2}, we obtain $A_0 R e^{2\lambda} = R (f_R)'$,
whence, for $R \ne 0$, it follows precisely the equality \rf{vv2}, from 
which \rf{vv1} is trivially obtained.

Second, an explicit form of \rf{ab-trace} is 
$e^{-2\lambda}(f_R)''= f - R f_R$. 
The l.h.s. of this equality can be rewritten by substituting $(f_R)''$ 
from \rf{vv1} (already proven) and further $(f_R)'$ from \rf{vv2}, 
giving $2A_0\lambda'$. For the r.h.s. of \rf{ab-trace} the same 
expression follows directly from \rf{eq-2o}. This completes the proof.  
Thus a solution of \rf{eq-2o} is automatically a solution of the whole 
set of vacuum field equations.

After solving \eqn{eq-2o}, it is necessary to substitute the solution to the integral 
condition \rf{mn-int}, which, using \eqref{eq-2o}, can be rewritten as
\beq                                        \label {mn-int'}
\int \e^{2\lambda} dy [f(R) - 4A_0 \lambda'] =0.
\eeq
since $\sqrt{g_2[y]} = \e^{2\lambda}$, and there is no $z$ dependence.
Thus the extra space geometry to be discussed below must satisfy the condition
\eqref{mn-int} or \eqref{mn-int'}. Furthermore, since we are seeking a solution with flat 
4D space-time, the condition $\Lambda_4 =0$ must be obtained for it automatically.

The order of \eqn{eq-2o} may be further lowered. Since its left-hand side is a function 
of $R$, say, $Q(R)$, assuming that it is monotonic, we can consider its inverse, 
$R = R(Q)$, and then \eqn{eq-2o} may be rewritten as (since $Q = 2A_0 \lambda'$)
\beq
R  = R(Q) = R (2A_0 \lambda') = -2\e^{-2\lambda} \lambda''.
\eeq
Next, let us use a standard trick and consider the new variable $v(\lambda) = \lambda'(y)$,
hence $\lambda'' = v\, dv/d\lambda$, and therefore,
\beq
-\e^{-2\lambda} v \frac {dv}{d\lambda} = R(2A_0 v)
\ \ \then \ \
\e^{2\lambda}d\lambda = - \frac{v\,dv}{R(2A_0 v)},
\eeq
an equation integrable by quadratures for an arbitrary (up to monotonicity 
of $Q(R)$) function $f(R)$.

\subsection{Two exceptional cases}

There are two exceptions: $\lambda' = \const$ and $A_0 =0$ which must be considered separately.

First, $\lambda' = \const$ leads to $R^y_y  = R^z_z = R=0$, see \rf{R_2}, that is, flat 2-space
(being conformally flat, its Riemann tensor is completely determined by the Ricci tensor components),
and we return to the flat toroidal version of extra dimensions discussed above. 

Second, $A_0=0$ leads to $f(R) = Rf_R$. The case $R = f_R =0$ is obviously unacceptable
(by \rf{Lam}, it gives $m_4=0$). Next, $R = R_0 = \const \ne 0$ means that the constants
$f(R_0)$ and $f_R(R_0)$ are either both zero or both nonzero. In any of these cases, according to \rf{Lam}, it is impossible to obtain simultaneously $m_4 > 0$ and $\Lambda_4 =0$.

It remains to suppose $R \ne \const$, then the equation $f(R) = Rf_R$ leads to 
$f(R) = \const\cdot R$ (which is 6D general relativity). The field equations \rf{ab} in this case 
read  $R^a_b - \half\delta^a_b R =0$ and are satisfied by an arbitrary function $\lambda (y)$ 
owing to the expressions \rf{R_2}. This total uncertainty in the shape of the extra space
cannot be regarded as a viable solution of our problem.

\subsection{Quadratic gravity}

Let us return to the general case $A_0\ne 0$, $\lambda'\ne\const$, 
and choose the function $f(R)$ in the form 
\beq                                 \label{f2}
f(R) = R^2 + bR + c
\eeq
then from \eq \eqref{eq-2o} one can get
\beq
R^2 = c - 2A_0\lambda',
\eeq
whence 
\beq
\pm \sqrt{c - 2A_0 \lambda'} = -2\e^{-2\lambda}\lambda''
\eeq
Substituting, as was suggested, $\lambda'  = v(\lambda)$, we obtain the relation
\beq
\pm\frac{1}{2}\e^{2\lambda}d\lambda = \frac{v\,dv}{\sqrt{c - 2A_0 v }}.
\eeq
Its integration results in
\beq
\frac{16}{9} \frac{c^3}{A_0^4} \left( 1-\frac{2A_0}{c} \lambda' \right) 
\left( \frac{A_0}{c} \lambda' + 1 \right)^2 = (\e^{2\lambda} + c_1)^2,
\eeq
where $c_1$ is an integration constant. With the notation 
$\chi := 2A_0\lambda/c$, this equation can be rewritten as
\beq                \label{chi'}
(1-\chi') (\chi' + 2)^2 = K (\e^{c\chi/A_0} + c_1)^2
\eeq
where $K = 9A_0^4/(8 c^3)$. Equation \rf{chi'} can be presented in the form 
\beq                 \label{chi}     
\chi = \psi (\chi') = \frac{A_0}{c}\ln\Big (-c_1 \pm (2+\chi')
\sqrt{(1-\chi')/K}\Big).
\eeq

Introducing the parameter $p=\chi'$, we can write the solution in 
a parametric form:
\bearr
\chi(p) = \psi(p),
\nnn           
y(p) = \int \frac{-3 A_0 dp}
{2c\big[(2+p)(1-p) \pm c_1\sqrt{K(1-p)}\big]}.
\ear          
This solution is rather cumbersome and inconvenient for analysis. 
Therefore, in what follows we will find solutions for $\lambda(y)$ numerically.

\subsection{Numerical results}

In this section, we will develop a numerical approach to make the results more convincing and evident. The metric \eqref{ds_6} is convenient for the above analytic calculations, but now,  to facilitate numerical simulations, we transform 
the extra-space metric to spherical-like coordinates:
\beq                                                 	\label{ds_7}
g_{ab} dy^a dy^b =  - e^{2\lambda}(dy^2 + dz^2) 
=  - \e^{2\mu (\theta)} [d\theta^2 + \sin^2\theta\, d\phi^2].
\eeq
where $\theta$ and $\phi$ are related to $y$ an $z$ by
\beq
z = \phi, \qquad y = \ln \tan (\theta/2), \qquad  
\e^\lambda = \e^\mu \sin \theta.
\eeq
The scalar curvature is expressed in terms of $\mu (\theta)$ as 
\beq \label{Ricci}
R = 2\e^{-2\mu}(1 - \mu_{\theta\theta} - \mu_\theta \cot \theta).
\eeq
where $\mu_\theta = d\mu/d\theta$ and 
$\mu_{\theta\theta} = d^2\mu/d\theta^2$.
The special case $\mu = \const$ corresponds to a 2D sphere with 
$R = \const = 2\e^{-2\mu}$. It has been already discussed in Section 3 
with a conclusion that it is incompatible with $m_4 >0$ and $\Lambda_4 =0$. 

We therefore assume $R \ne \const$ and return to \eqn{vv2} which now reads 
\beq
(f_R)' = A_0 e^{2\lambda} \equiv A_0 e^{2\mu}\sin^2\theta.
\eeq
Under the assumption \eqref{f2}, it becomes (since $d\theta/dy = \sin\theta$)
\beq \label{eq-3o}
R_\theta =\frac{A_0}{2} e^{2\mu}\sin(\theta),
\eeq
it is a third-order linear equation with respect to $\mu(\theta)$. 
To solve it, we need three boundary conditions, which we impose
at $\theta=0$: $\mu(0)=0$, $\mu_\theta(0)$ and $R(0)$. The constant $R(0)$ 
is connected with the previously introduced integration constant $A_0$.
Indeed, from \eqref{eq-2o} it follows
\beq          \label{eq-2o2}
f(R) - R f_R  = 2A_0 (\mu_\theta\sin\theta +\cos\theta),                 
\eeq
which, taken at $\theta =0$, gives the relation 
\beq            \label{A0}
f(R(0))-R(0)f_R (R(0)) = 2A_0 .
\eeq
Equation \rf{eq-2o2} is a nonlinear second-order differential equation 
with respect to $\mu$, and \eqn{eq-3o}, being linear, is more 
convenient for numerical integration. 

The solution must satisfy the integral condition \eqref{mn-int}, which can be rewritten in the form
\beq                                        \label {mn-int1}
\int\sqrt{g_2[y]} d^2 y[f(R) + 4A_0(\mu_\theta \sin\theta +\cos\theta)] =0.
\eeq
This condition follows from \eqref{mn-int0}, where the second term
is total derivative, so that
\beq           \label{Lameq}
\int d\theta \Box f_R = \sin\theta \cdot (f_R)_{\theta}\Big|_0^{\pi}=0.
\eeq
The second equality holds if the Ricci scalar behavior is not too pathological. 
It means that the conditions \eqref{Lam} and \eqref{mn-int0}  
coincide provided $\Lambda_4 =0$. We have verified this statement numerically.

We have solved \eqn {eq-3o} numerically with \eqref{Ricci} under some
boundary conditions for $\mu(\theta)$ and $R(\theta)$ (see Fig.\,1). 
By varying these initial parameters one can obtain a very small or 
even zero $\Lambda_4$ and thus make it possible to have a
flat 4D space-time (see Fig. 2). 
In this case, the integral condition \rf {mn-int1} is satisfied.

\begin{figure}[h]
	\begin{minipage}[h]{0.32\linewidth}		
		\center{\includegraphics[width=1.\linewidth]{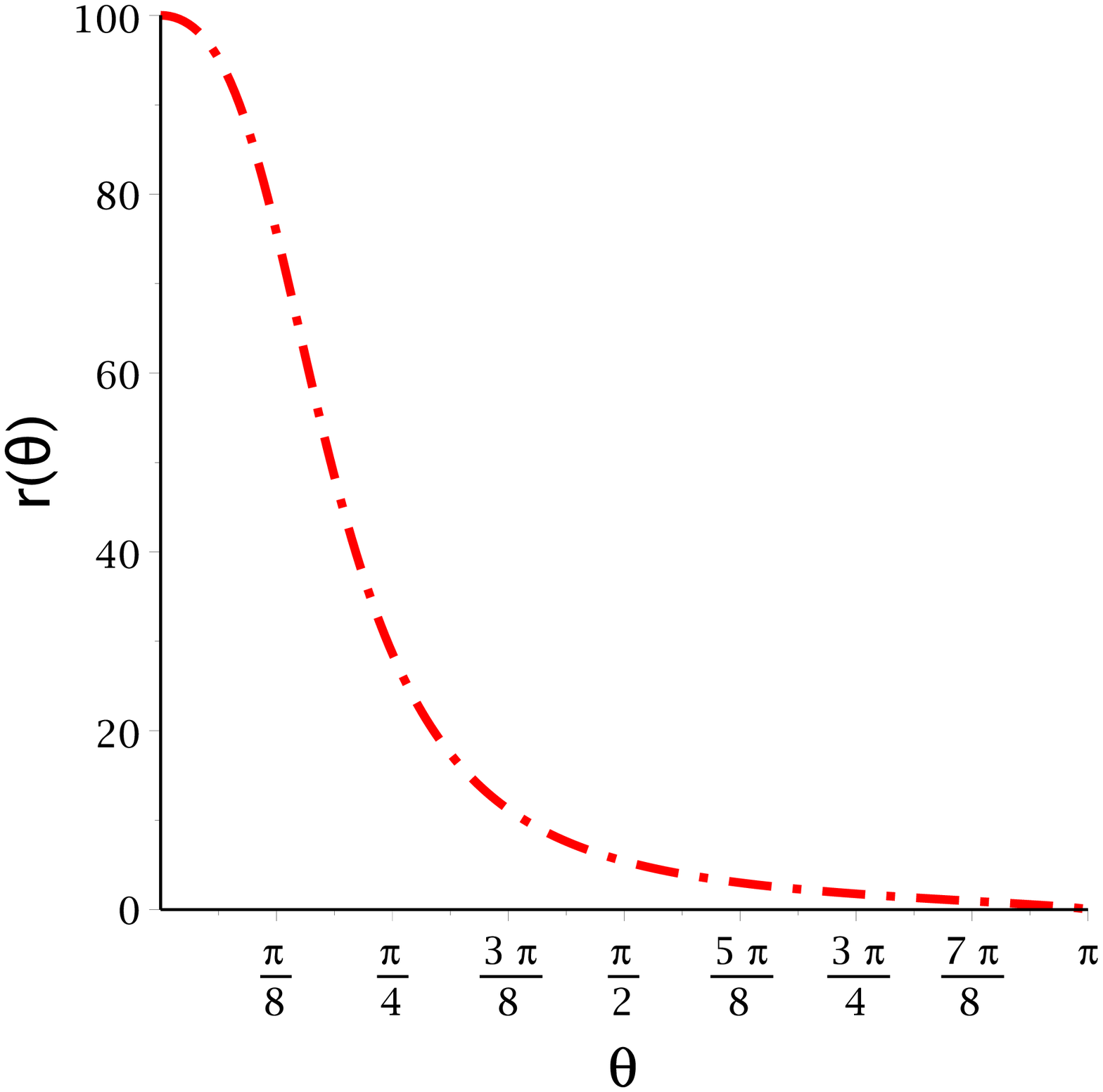}}
	\end{minipage}
	\hfill
	\begin{minipage}[h]{0.34\linewidth}
		\center{\includegraphics[width=1.4\linewidth]{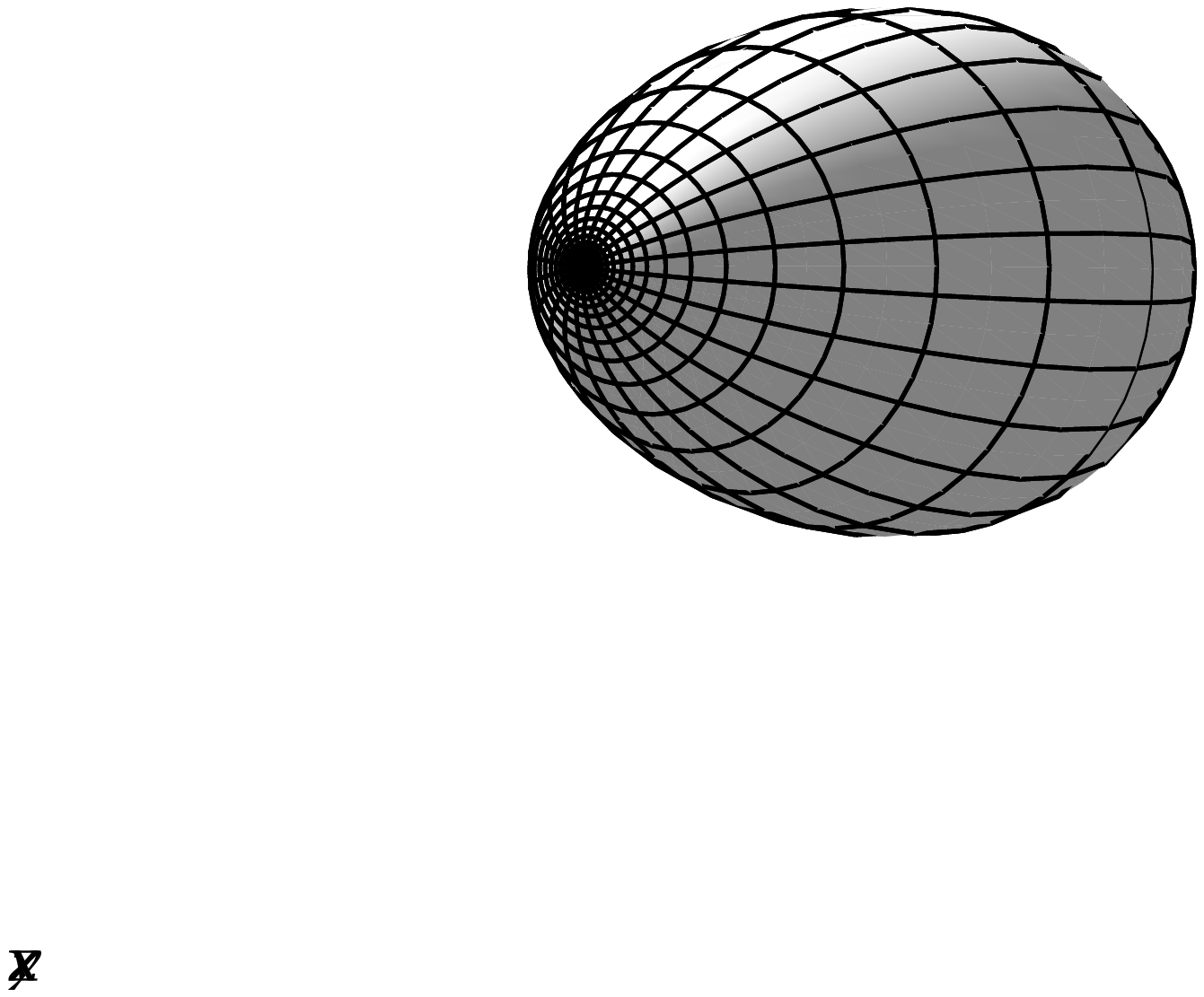}}
	\end{minipage}
	\hfill
	\begin{minipage}[h]{0.32\linewidth}
		\center{\includegraphics[width=0.8\linewidth]{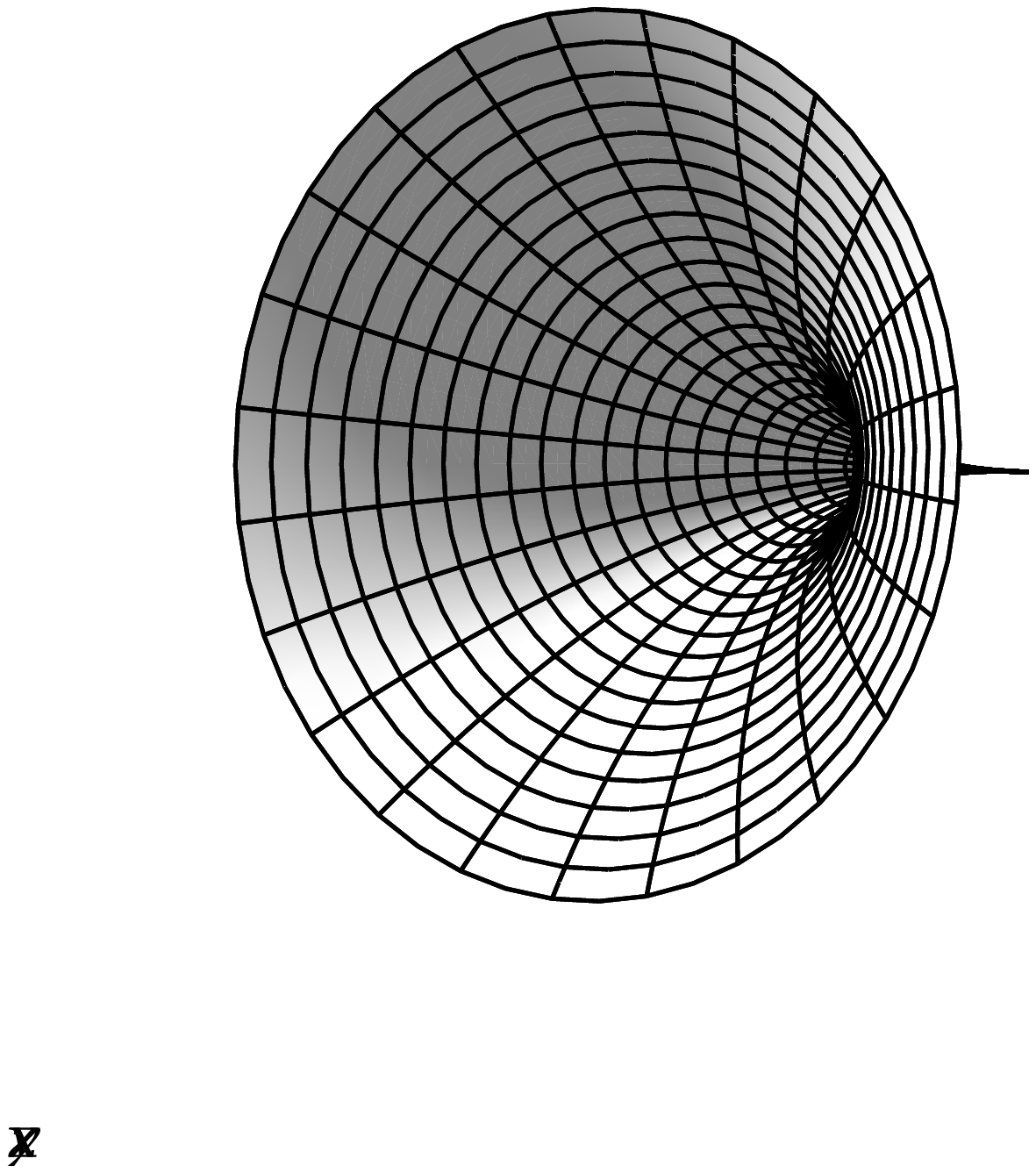}}
	\end{minipage}
	\\
	\hspace*{3cm} (a) \hspace{5.5cm} (b)  \hspace{4cm} (c)  
	\caption{
		(a) Radius $r(\theta)=e^{2\mu}$ in extra space for the parameter  
		values $b=-10^{-2}$, $c = 2.1 \cdot 10^{-5}$,  $\mu_\theta(0) =0, 
		\mu(0) = \ln (100)$, $R(0) = 1.01 \cdot 10^{-3}$.  
		(b) 3D plot of the solution. (c) A small part of the  3D plot near   $\theta=\pi$ (left ``end'').
	}
	\label{ris:image1}
\end{figure}
As a result, we have proved that a compact extra space with a 
deformed metric is compatible with 4D Minkowski space. In the next 
section we consider the next approximation in $\ep$.
.
\begin{figure}[h!]
	\center{\includegraphics[width=0.5\linewidth]{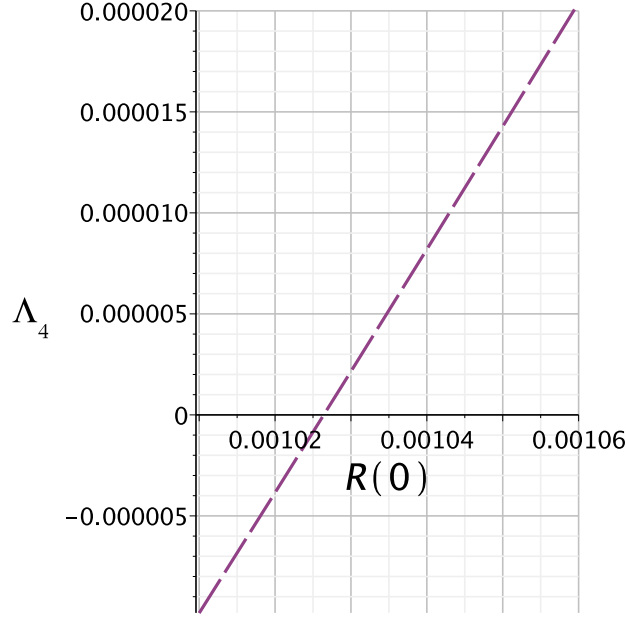} }
	\caption{ 
		The cosmological constant $\Lambda_4$ as a function of the boundary condition $R(0)$.  
		The other parameter values are $b=-10^{-2}$, $c = 2.1 \cdot 10^{-5}$, $\mu_\theta(0) =0$,  
		$\mu(0) = \ln (100)$. It is seen that $\Lambda_4=0$ at $R(0)\simeq 1.025\cdot 10^{-3}$. }
	\label{ris:image1}
\end{figure}
\section{First-order approximation. Stability of the deformed extra space}

Let us study the stability of the solution obtained above under small 
$x$-dependent perturbations. We suppose that the 4D metric is flat, 
as before (since its perturbations should be of the order $O(\ep^2)$), 
and consider the perturbed extra-space metric in the form
\beq     	\label{ds_8}
g_{ab} dy^a dy^b = 
=- \e^{2\mu (\theta)+2\beta (x)} [d\theta^2 + \sin^2\theta\, d\phi^2].
\eeq
Then, after integrating out the extra dimensions in the action, 
$\beta(x)$ behaves as a scalar field in 4D Minkowski space-time
\cite{BlackHoles}. In the previous sections $\beta = 0$.

The field $\beta$ can in general have complicated kinetic and potential 
terms. If the potential has a minimum, and $\beta$ settles at this minimum, 
then such a background configuration with its extra-space geometry
is stable.


Consider the case $\beta(x) \ll 1$ with the extra metric obtained in  
\eqref{ds1} and \eqref{ds_7}. A back reaction of $\beta$ on the extra 
metric is neglected. As before,
\beq
f(R) = R^2 +bR +c,~ \cm  
S = \frac{m_D^4}{2}\int d^4x d^2y \sqrt{g_4g_2} f(R).
\eeq
In general, the action contains three parameters $b$, $c$ and $m_D$. 
For the 2D extra metric me have the expressions
for the determinant and the Ricci scalar:
\beq 		\label{key}
\sqrt{g_2} = e^{2\beta (x)}r^2(\theta)\sin \theta,\qquad 
R_2 = e^{-2\beta (x)}\bar{R}(\theta),
\eeq
The 6D Ricci scalar has the form
\beq                      \label{2+4}
R = R_2 + R_4 + 4 \Box \beta + 6 (\d\beta)^2 + o(\ep^2), 
\eeq
where $\Box \beta$ and $(\d\beta)^2$ are the 4D d'Alembertian and 
squared gradient of $\beta(x)$. Using the expansion \rf{act2} for the 
action $S$ with $R$ given by \rf{2+4} and $R_2$ given by \rf{key},
integrating out the extra dimensions and excluding a total divergence 
related to $\Box \beta$, we obtain
\beq \label{act4}
S\simeq \frac{m_6^4}{2} \int d^4 x \sqrt{g_4}
\left[ h(\beta) R_4 +  (6 - 4h_\beta)(\d \beta)^2 - 2W(\beta)\right],
\eeq
where $h_\beta \equiv dh/d\beta$, the function $h(\beta)$ and the potential 
$W(\beta)$ having the form 
\bearr                  \label{def-h}
h(\beta(x)) = 2I_1 + e^{2\beta(x)}bI_0,
\yyy                    \label{W}
W(\beta(x)) = -\frac{1}{2}\left(I_2 e^{-2\beta(x)} 
+ cI_0 e^{2\beta(x)} + bI_1\right),
\ear
and $I_k$ are the integrals
\beq
I_k = 2\pi \int d\theta r^2(\theta)\sin \theta\, \bar{R}(\theta)^k,
\qquad k=0,1,2.
\eeq
We have restored the 6D Planck mass $m_6$, which was equal to unity 
in the previous sections.  
The parameters $I_k$ depend on the metric $g_{ab}(y)$. The latter is assumed
to be one of the solutions to the field equations discussed in the previous
section.

The action \rf{act4} is the vacuum action of a 4D scalar-tensor theory of
gravity  written in its Jordan frame.  The conformal mapping
\beq
\tilde{g}_{\mu\nu} = h(\beta)g_{\mu\nu}
\eeq 
transforms it to the Einstein frame. The result is  
\beq                   \label{SE}
S = S_E = \frac{m_6^4}{2} \int d^4 x \sqrt{\tilde g_4} (\sign h)
\left[ \tilde R_4 + K(\beta) ({\tilde \d}\beta)^2 - 2U(\beta)\right],
\eeq 
where the tilde denotes quantities obtained from or with ${\tilde g}\mn$, and
\beq 
K(\beta) = \frac{6 - 4 h_\beta}{h} + \frac 32 \frac{h_\beta^2}{h^2},
\cm
U(\beta) = (\sign h) \frac{W(\beta)}{h^2(\beta)}. 
\eeq 
More explicitly,
\bearr
K(\beta) = \frac{2}{h^2}\Big[6 I_1 + (3 - 8 I_1) bI_0 \e^{2\beta} 
- 2b^2 I_0^2 \e^{4\beta}  \Big],
\nnn 
U(\beta) = - \frac{(\sign h)}{2 h^2}
\Big(I_2 \e^{-2\beta} + b I_1 + c I_0 \e^{2\beta} \Big). 
\ear 

Evidently, $\beta =0$ should correspond to our ``background'' solution from
the previous section, with 4D Minkowski space and an inhomogeneous 2D 
extra space geometry. For this solution to be in stable equilibrium under 
the emergence of the 4D scalar $\beta(x)$, we should have
\beq                     \label{min-U}
U(0) = 0,\qquad  U_\beta(0) = 0,\qquad U_{\beta\beta}(0) > 0,
\eeq  
provided that $K(0) > 0$. Let us make sure that this is really the case for the parameter
values used in Figs.\,1 and 2, that is, $b = -10^{-2}$ and $c = 2.1\ten{-5}$. 

To this end, it is useful to recall that for solutions with flat 4D space
(that is, for $\beta = 0$), as follows from \eqn {int-0}, we have 
\beq                       \label{III}
I_2 = cI_0, \qquad \ bI_1 = -2I_2, \ \ \then\ I_1 = -2c I_0/b. 
\eeq
Then, for the potential at small $\beta$ we have 
\beq                \label{UE} 
U(\beta) =  \frac{\sign h}{2 h^2} c I_0\ [\cosh (2\beta) -1],
\eeq
which evidently satisfies the conditions \rf{min-U} if $c \sign h > 0$
since $I_0>0$. This is a general result in quadratic 6D $f(R)$ gravity \rf{f2}
for any model from the class under consideration.   

Next, for the function $h(\beta)$ and the kinetic term at $\beta =0$ we have 
due to \rf{III}
\beq 
h(0) = \frac{I_0}{b}(b^2 - 4c), \qquad
K(0) = \frac{2}{(b^2-4c)^2}\Big(5b^4 -8 b^2 c + 48 b^4\Big).
\eeq
Substituting the values $b = -10^{-2}$ and $c = 2.1\ten{-5}$, we obtain 
\beq 
h(0) = -1.6\ten{-3}, \qquad  K(0) \approx 425.
\eeq 
Thus we have $h(0)< 0$ (which is unimportant) and $K >0$, so that, according 
to \rf{UE}, our model is stable under slow $x$-dependent changes in the extra-space metric.

The parameter values were selected so that the final expression for the action \eqref{SE} contains proper signs of the kinetic term $K$ and the potential $V$, as follows from Figure 3.
The common sign of the total action does not affect both the classical behavior and quantum transitions, see a discussion in \cite{BlackHoles}. Indeed, the transition amplitude is expressed in the path integral technique as $\int\exp{(iS[q])}Dq$ where $q(t)$ is some dynamic variable. The transition probability $\int\exp{ (iS[q_1]- iS[q_2])}Dq_1Dq_2$ 
is invariant under the substitution $i \to -i$ (with interchanging the integration
variables $q_1$ and $q_2$ ).

\begin{figure}[ht]
	\centering
	\includegraphics[width=0.5\linewidth]{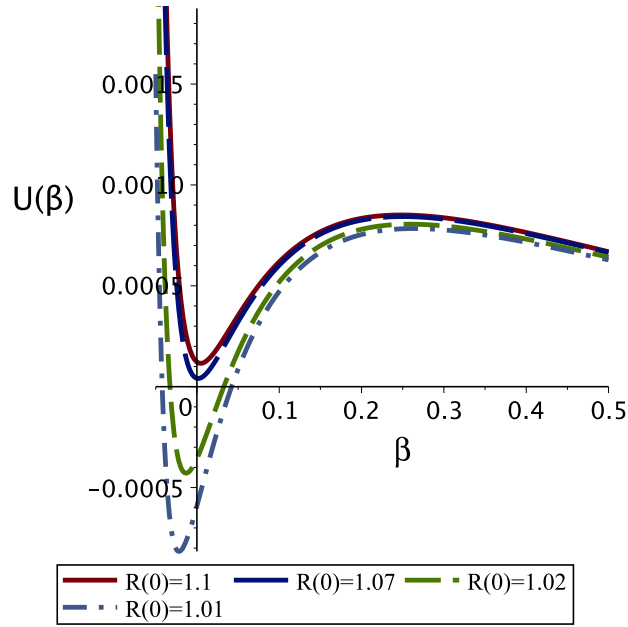}
	\caption{The potential $U(\beta)$ for the parameter values $b = -10^{-2}$, 
		$c = 2.1\cdot10^{-5}$, $\mu(0) = \ln(100)$.}   
	\bigskip   
	\label{ris:image1}
\end{figure}

Specifying the boundary values $R(0)$ and $\mu(0)$ (equivalently, $r(0)$)
and calculating the parameter $A_0$ using \rf{A0}, we can solve \eqs 
(58) and (59) with a given function $f(R)$, in particular, \rf{f2} 
with its parameters $b$ and $c$. The potential $U(\beta)$ for a certain 
range of $\beta$, the same parameter values and different boundary values 
$R(0)$ is presented in Fig.\,\ref{ris:image1}.
One can see that the behavior of $U(\beta)$ confirms the existence of a 
minimum with $U=0$ at $\beta =0$ for $R(0)\approx 1.0025$, which corresponds 
to our model obtained in Section 4, with $\Lambda_4 = 0$ and flat 4D space. 

Returning to the action \rf{act4} with $\beta =0$, we can identify the 
constant coefficient of $R_4$ with $m_4^2/2$, where $m_4$ is the 4D Planck mass
related to the Newtonian gravitational constant $G_N$:
\beq 
m_4^2 = (8\pi G_N)^{-1} = m_6^4 |h(0)| = m_6^4 I_0 |b-4c/b|\simeq 16m_6^2,
\eeq 
where the integral $I_0 =8\cdot 10^{3}/m_6^2$ according to the numerical
simulations. As a result, the $D$-dimensional Planck mass is only a few 
times smaller than the 4D one.

\section{Conclusion}

We have shown that an inhomogeneous (deformed) compact extra space may 
be considered as a promising tool for physics beyond the Standard Model. 
Such spaces satisfy the necessary conditions: (i) the deformed extra space 
does not contradict the observable value of the cosmological constant, 
(ii) the extra dimensions are stable at least relative to the ``radion mode''. 
The latter represents a scalar field in 4D space. To be considered as the 
inflaton, this mode should have a mass of order $10^{13}$ GeV, which needs 
fine tuning since natural values of this mass are sub-Planckian.

We have also proved that maximally symmetric extra spaces prevent the 
existence of solutions with 4D Minkowski space or de Sitter space with 
a cosmological constant compatible with observations.

The analysis was performed on the basis of pure gravity without invoking any additional
ingredients. This lays a foundation for further research based on inhomogeneous extra 
spaces.

\subsection*{Acknowledgments}

This work was supported by Russian Science Foundation and fulfilled in the framework of MEPhI Academic Excellence Project (contract № 02.a03.21.0005, 27.08.2013) and according to the Russian Government Program of Competitive Growth of Kazan Federal University. The
work of S.G.R. was also supported by the Ministry of Education and Science of the Russian Federation, Project № 3.4970.2017/BY. The research of K.B. was supported within the RUDN-University program 5-100 and by RFBR grant 16-02-00602.

\end{document}